\documentclass[12pt]{article}
\usepackage[margin=0.7in]{geometry}
\usepackage[font=small,labelfont=bf]{caption}
\usepackage[numbers,square,sort&compress]{natbib}
\usepackage{authblk}
\usepackage{graphicx, amsmath}
\usepackage[latin1]{inputenc}
\usepackage[dvipsnames]{xcolor}

\makeatletter \renewcommand{\@biblabel}[1]{#1.} \makeatother
\graphicspath{%
{figs/}%
}

\begin{document}

\title{Quantum Gravity Corrections to the Mean Field Theory of  Nucleons}

\author[1]{Abrar Ahmed Naqash}
\author[2]{Barun Majumder\footnote{Corresponding author email: barun@utk.edu}}
\author[3]{Soumodeep Mitra}
\author[4]{Moomin Mushtaq Bangle}
\author[5,6,7]{Mir Faizal}

\affil[1]{Department of Physics, University of Kashmir, Srinagar, 190006 India}
\affil[2]{University of Tennessee, Knoxville, TN  37916 USA}
\affil[3]{School of Physical Sciences, Indian Association for the Cultivation of Science, Kolkata, 700032 India}

\affil[4]{Department of Physics, Cluster University Srinagar,  Srinagar, 190001 India}

\affil[5]{Canadian Quantum Research Center, 204-3002, 32 Ave Vernon, BC V1T 2L7 Canada}
\affil[6]{Department of Physics and Astronomy, University of Lethbridge, Lethbridge, AB T1K 3M4, Canada}
\affil[7]{Irving K. Barber School of Arts and Sciences, University of British Columbia Okanagan Campus, Kelowna,  BC V1V 1V7, Canada}
 \date{}
 \maketitle

\begin{abstract}
 In this paper, we analyze the correction to the  mean field theory  potential for a system of  nucleons. It will be argued that these corrections can be obtained by deforming the Schr\"{o}dinger's equation describing a  system of nucleons by  a minimal length in the background geometry of space-time. This is because such  a minimal length occurs due to quantum gravitational effects, and modifies the low energy quantum mechanical systems. In fact, as the  mean field potential for the nucleons is represented by the  Woods-Saxon potential, we will explicitly analyze such corrections to this potential. We will  obtain the corrections to the energy eigenvalues of the deformed Schr\"{o}dinger's equation for the Woods-Saxon potential. We will also construct the wave function for the deformed Schr\"{o}dinger's equation. 
\vspace{0.5cm}
\end{abstract}

\section{Introduction}
At present  there are various different   approaches to quantum gravity.
It is important to know the universal features of all these different approaches to quantum gravity. It is expected from almost all of these approaches that the  geometric  of space-time has an intrinsic minimal length associated with it  \cite{s16}. Any quantum theory of gravity, should be consistent with black hole physics. Now it is known that we need higher energies to probe smaller length scales. The energy  to probe Planckian region of space is equal to  the energy needed to form a black hole in that region of space. Thus, if we try to make Planckian measurements, we will form mini black holes, which will prevent us from making such  measurements   \cite{z4,z5}. So, it is not possible to probe space-time below Planck scale, and Planck length acts as an intrinsic minimal length in space-time.  
Such a minimal length exists even in asymptotically safe gravity \cite{asgr} and conformally quantized quantum gravity \cite{cqqg}. Such a minimal length in the background geometry of space-time also exists in loop quantum gravity \cite{z1, po12, po14, 12p} because of polymer quantization, as the polymer length acts as a minimal length in loop quantum gravity.

It can be argued that there exists a minimal length of the order of Planck length in the background geometry of space-time \cite{s16}. However, it is possible for this minimal length to be of several order of magnitude larger than the Planck length. This is because in string theory it is possible to relate this minimal length to the string length \cite{z2,zasaqsw}. In fact, it can be  demonstrated that  this  minimal measurable length $l_{min}$ can be directly related to string length $l_s = \alpha'$ as $l_{min} = g_s^{1/4} l_s$, with $g_s$ as the string coupling constant \cite{cscds,2z}. Even though it is possible to obtain point like $D0$-branes due to  non-perturbative effects, it has been demonstrated that there is a minimal length even  with such  non-perturbative point like objects. The minimal length in string theory, with such  non-perturbative objects is related to the string length as $l_{min} = l_s g_s^{1/3}$ \cite{s16, s18}. So, by adjusting the string coupling constant, the string length can be several orders larger than the Planck length. 

Another reason is that string theory is invariant under T-duality. So, for a string compactified on a circle with radius $R$, the mass states do not change  under the transformation  $R\to l_s^2/R$, and $n\to w$, where $n$ are the Kaluza-Klein modes and $w$ are the winding modes. Thus, it is not possible to obtain new information by going below a zero point length in  string theory compactified on a compact geometry \cite{s16}. It is also possible to construct an effective path integral for the center of mass of the strings compactified on a circle,  by neglecting all the string  oscillation modes. This effective path integral can be used to obtain suitable propagators  in such a theory, and it can be explicitly demonstrated that such propagators are invariant under T-duality \cite{green1, green2}.
Thus, there is an intrinsic zero point  length,  (larger than the Planck length) associated  with such propagators, even if the string oscillation modes are neglected. It may be noted as the double field theory is constructed using the T-duality \cite{df12, df14}, it is expected that a such zero point length can  also occur  in the double field theory. It has been demonstrated that such a zero point length  in double field theory can have important consequences for the generalized geometries used in string theory \cite{mi15}. It may be noted that such a minimal length larger than the Planck length can make it possible for the mini  black holes to form in particle physics colliders (due to the lowering of Planck scale) \cite{co12, co14}.  Thus, the existence of such a minimal length, larger than the Planck length, can have interesting physical consequences. 

It is possible for such a minimal length to deform the low energy quantum mechanical systems  \cite{ml12, ml21}. This deformation of such low energy quantum mechanical systems can detected by ultra precise measurements of Landau levels and Lamb shift in atomic systems \cite{ml15}. It has also been proposed that such a deformation of low energy quantum mechanics can be detected using an opto-mechanical setup \cite{ml14}.  It is possible to use a  gravitational spectrometer  to measure the interaction between neutrons and a gravitational field 
\cite{ne12ab, ne14ab}. As this system will also be deformed by the deformation of quantum mechanics from a minimal length, it is possible to use  a gravitational spectrometer to measure the effects of minimal length on such a system  \cite{ml16}.   Thus, the deformation of the  Heisenberg algebra from minimal measurable length can have interesting low energy consequences.

It may be noted that it is possible to study several different forms of deformation of quantum mechanics \cite{qft2, qft4, qft5, qft6, qft7}. We will use a specific form of  deformation \cite{line12, line14, bb1,bb2,bb3,bb4,bb5,bb6,bb7,line16, line18}, which is consistent with the effects produced from  various  different theories, such as non-locality \cite{nonl}, doubly special relativity \cite{dsr}, deformed dispersion relations in the bosonic string theory \cite{stri}, Horava-Lifshitz gravity \cite{HoravaPRD, HoravaPRL}, discrete space-time \cite{Hooft}, models based on string 
field theory \cite{Samuel1}, space-time foam \cite{Ellis}, spin-network \cite{Gambini}, 
and noncommutative geometry \cite{Carroll}. 
We will study the effects of such a deformation in the mean field theory  potential for a system of  nucleons. Such a mean field theory potential for a system of  nucleons can be described by a  Woods-Saxon potential \cite{wood, wood1, wood2, wood4, wood5}.  It may be noted that the correction to the Woods-Saxon potential from the simplest deformation of the Heisenberg algebra has been studied \cite{dwood, dwood1, dwood2, dwood4, dwood5}. 
However, here we will deform the Schr\"{o}dinger's equation with this  potential using the  deformation of the Heisenberg algebra, with linear terms  \cite{line12, line14, bb1,bb2,bb3,bb4,bb5,bb6,bb7, line16, line18}. It may be noted that the deformation of the angular momentum algebra consistent with this algebra has been studied \cite{an12, an14}, and we will use this deformation of the angular momentum algebra to analyze the deformed Woods-Saxon potential. 


\section{ Deformed Woods-Saxon Potential}
\begin{figure}
\begin{tabular}{c}
\includegraphics[width=.45\linewidth]{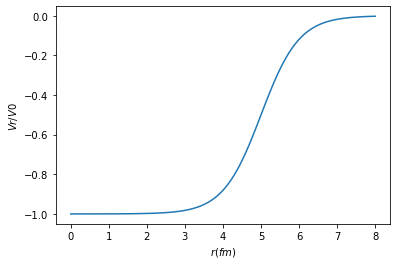} \hspace*{.2cm}\includegraphics[width=.45\linewidth]{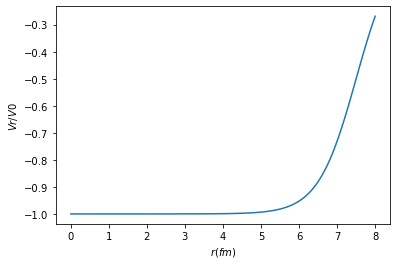}
\end{tabular}
	
	\caption{Woods Saxon potential for two  different radius. We have taken the parameters $a = 0.5 fm$ and the radius $a_0$ to be $5fm$ for the left and $7fm$ for the right plot. }\label{WS_plot}
\end{figure}

 In this section, we will analyze the corrections to the  mean field theory  potential for a system of  nucleons from a deformation of the Heisenberg algebra. As the  mean field theory  potential for a system  nucleons is represented by the  Woods-Saxon potential \cite{wood, wood1, wood2, wood4, wood5}, we will analyze such corrections to the Woods-Saxon potential. Even though there are various different models of nuclear potential,  the advantage of using Woods-Saxon potential is that it explain a larger set of magic numbers \cite{wood, wood1, wood2, wood4, wood5}. Furthermore, this potential has been used to explain a large number of nuclear systems. The  Wood-Saxon nuclear potential can explain the effect of  the chiral magnetic field in relativistic heavy-ion collisions \cite{k1}. 
 The prolate-shape predominance of the nuclear ground-state deformation  for two thousand   nuclei 
 has been obtained using this potential \cite{k4}. 
 This potential has been used to investigate the toroidal and superdeformed configurations in light atomic nuclei  \cite{k5}. It has also been used to explain the hyperfine structure of heavy ions \cite{k6}. As the Wood-Saxon nuclear potential has been used to investigate a large number of nuclear systems,  it is possible to test its deformation using such nuclear systems. So, we  will analyze its corrections from the deformation of the Heisenberg algebra. The  Woods-Saxon  potential can be expressed as 
\cite{wood, wood1, wood2, wood4, wood5}
\begin{equation}\label{woods-saxon}
	V_{WS} (r) = - \frac{V_0}{1+ exp(\frac{r-a_0}{a})}, ~ a<<a_0
\end{equation}
Here  $a_0$ denotes radius of nucleus,  which in turn is connected to the mass number by $a_0  = c A^{1/3}$ with $c=1.25 fm$. The parameter $a$ characterize the thickness of superficial layer.  Furthermore, $V_0$   represents the depth of the  potential well.  
We have plotted  the form of this potential in Fig. \ref{WS_plot}. We plotted the potential for two  different radius, corresponding to two  different mass numbers. The potential is similar to infinite potential well, except that the  walls are smooth. The plot on the left is drawn for a nucleus with mass number $A = 125$ and the plot on the right has been drawn for $A = 216$. It may be observed that  with increasing mass number,  the potential resembles that of a potential of a potential well. 
 
 Now we  will analyze the  correction to this  Woods-Saxon potential from a deformation of the Heisenberg algebra with a linear term   \cite{line12, line14, bb1,bb2,bb3,bb4,bb5,bb6,bb7,line16, line18}. It may be noted that this deformation is consistent with various  different theories, such as non-locality \cite{nonl}, doubly special relativity \cite{dsr}, deformed 
dispersion relations in the bosonic string theory \cite{stri}, Horava-Lifshitz gravity \cite{HoravaPRD, HoravaPRL}, discrete space-time \cite{Hooft}, models based on string field theory \cite{Samuel1}, space-time foam \cite{Ellis}, spin-network \cite{Gambini}, and noncommutative geometry \cite{Carroll}. 
This deformation can be explicitly written as  
 \cite{line12, line14, bb1,bb2,bb3,bb4,bb5,bb6,bb7, line16, line18}
\begin{equation} \label{commutation}
	[x_i, p_j] = i\hbar \left[\delta_{ij} -\alpha \left( p\delta_{ij} + \frac{p_ip_j}{p}\right)  + \beta^2\left(p^2 \delta_{ij} + 3 p_ip_j\right)        \right]
\end{equation}      
Now  this deformation of the Heisenberg  algebra will produce  a deformation in the coordinate representation of the momentum operator from $p$ to $p \left(1 - \langle\mathcal{C}\rangle\right)$, where  
\begin{equation}
	\langle\mathcal{C}\rangle = \alpha \langle p\rangle - \beta^2 \langle p\rangle^2
\end{equation}
It is important to define a coefficient   $\lambda$ as 
\begin{equation}
  \lambda  = \left(1 - \langle\mathcal{C}\rangle\right)
\end{equation} 
It may be noted here that for $\alpha \langle p\rangle > \beta^2 \langle p\rangle^2 $, $	\langle\mathcal{C}\rangle  > 0$, and $\lambda < 1$.  However,  for $\alpha \langle p\rangle < \beta^2 \langle p\rangle^2 $, $	\langle\mathcal{C}\rangle  <  0$, and $\lambda >1 $. So,  when  $\lambda<1$,  the linear term dominates and when  $\lambda >1$, the quadratic term dominates. The corrections vanish at $\lambda = 1$, which corresponds to $	\langle\mathcal{C}\rangle =0$.  It may be noted that the corrections to the Woods-Saxon potential from  such a  quadratic term have already been analyzed \cite{dwood, dwood1, dwood2, dwood4, dwood5}.  However, here we will analyze it for both linear and quadratic terms. The deformation of the  momentum operator by  $\langle\mathcal{C}\rangle$  in turn deforms the angular momentum algebra for this system  as \cite{an12, an14}
\begin{align}
L^2 |plm\rangle = \hbar^2 l(l+1) \left(1 - \langle\mathcal{C}\rangle\right)^2 |plm\rangle ~ \label{ang_mom1}
\\
L_z |plm\rangle = \hbar m \left(1 - \langle\mathcal{C}\rangle\right) |plm\rangle ~\label{ang_mom2}
\end{align} 
We will use these modified angular momentum operators  to explicitly analyze the corrections to the Woods-Saxon potential. It may be noted that the original radial Schr\"{o}dinger's equation   with Woods-Saxon potential is given by  \cite{wood, wood1, wood2, wood4, wood5},
\begin{equation} \label{SE}
	\frac{d^2 \mathcal{R}}{dr^2} + \frac{2}{r} \frac{d\mathcal{R}}{dr} + \frac{2m}{\hbar^2} \left[E - V_{WS} -\frac{\hbar^2 l(l+1)}{2mr^2} \right] \mathcal{R} = 0
\end{equation}
Now we can write the deformation of this Schr\"{o}dinger's equation   from a deformation of the Heisenberg algebra as \cite{an12, an14} 
\begin{equation} \label{GUP_SE1}
	\left(1 - \langle\mathcal{C}\rangle\right)^2 \left[\frac{d^2 \mathcal{R}}{dr^2} + \frac{2}{r} \frac{d\mathcal{R}}{dr}\right] + \frac{2m}{\hbar^2} \left[E - V_{WS} -\frac{\hbar^2 l(l+1)}{2mr^2} \left(1 - \langle\mathcal{C}\rangle\right)^2 \right] \mathcal{R} =0
\end{equation}

\section{Deformed Schr\"{o}dinger's Equation}
We will use   Nikiforov-Uvarov method \cite{nu12, nu14}  to solve the deformed Schr\"{o}dinger's equation for the Woods-Saxon (see Appendix A).   We start with  $R = {U(r)}/{r}$,   we can write the deformed 
 Schr\"{o}dinger's equation  with Woods-Saxon potential as  
\begin{equation} \label{GUP_SE2}
\frac{d^2 U}{dr^2} + \frac{2m}{\hbar^2 \lambda^2} \left[E - V_{WS}(r) - V_l(r) \right] U = 0 
\end{equation}  
Here we have expressed the   total  effective potential as a sum of the original Woods-Saxon potential $V_{WS} (r)$ and the deformation of the potential  $V_l (r)$ as 
\begin{eqnarray} \label{eff_V}
	V_{eff} = V_l(r) + V_{WS}, && 
	 V_l(r) = \frac{\hbar^2 l(l+1)}{2mr^2} \lambda^2
\nonumber 	\\
	V_{WS} = \frac{V_0}{1+ exp(\frac{r-a_0}{a})}. && 
\end{eqnarray}
Now with the   two new variables  
$	x = (r- a_0)/a_0,  $ and $ ~ \alpha = {a_0}/{a}$, and we can simplify   the effective potential using   Pekeris approximation (see Appendix B). 
The deformed  Schr\"{o}dinger's equation, can be written as a ordinary Schr\"{o}dinger's equation, where the Woods-Saxon potential has been replaced by this   effective potential
\begin{equation} \label{GUP_SE3}
	\frac{d^2 U}{dr^2} + \frac{2m}{\lambda^2 \hbar^2} \left[ E - \Delta_\lambda A_0 + \frac{V_0 - \Delta_\lambda A_1}{1+ exp(\frac{r-a_0}{a})} - \frac{\Delta_\lambda A_2}{[1+ exp(\frac{r-a_0}{a})]^2} \right]
\end{equation}
We can solve the  deformed Schr\"{o}dinger's equation   by   Nikiforov-Uvarov method  \cite{nu12, nu14}   by  introducing the dimensionless quantities
\begin{eqnarray} \label{para}
	\epsilon^2 (\lambda) =  -\frac{2m a^2 (E - \Delta_\lambda A_0)}{\lambda^2 \hbar^2}, &&  \beta^2(\lambda) = \frac{2m a^2 (V_0 - \Delta_\lambda A_1)}{\lambda^2 \hbar^2}, \nonumber  \\  \qquad \gamma^2  =  \frac{2m a^2 (\Delta_\lambda A_2)}{\lambda^2 \hbar^2}. && 
\end{eqnarray}
The term   $\Delta_\lambda$ scales as $ \lambda^2$, making $\gamma$ which scales  as $\Delta_\lambda/ \lambda^2$ independent of $\lambda$.    So,  using  a new variable 
$	z = [ 1+ \exp (({r-a_0}){a}^{-1})]^{-1}
$, we can write the deformed  Schr\"{o}dinger's equation as  
\begin{equation}\label{GUP_SE4}
	\frac{d^2 U(z)}{dz^2} + \frac{1-2z}{z(1-z)} \frac{dU(z)}{dz} + \frac{-\epsilon^2 (\lambda) + \beta^2(\lambda) z  - \gamma^2 z^2}{[z(1-z)]^2}  U(z) = 0
\end{equation}
We can explicitly obtain the solution to this deformed  Schr\"{o}dinger's equation (see Appendix C).  We can also obtain   condition for the bound states to exist in this deformed Woods-Saxon potential as 
$
-n^{'2}<[\beta^2(\lambda)-\gamma^2]<n^{'2}, 
$ with 
$n^{'}= -n + {({1+4\gamma^2})^{1/2}-1}/{2}
$. 
The bound states only exist for $n^{'} > 0$. Using the value of $\gamma $, we can 
 write this bound as $0 \le 2 n\le  (1+{192a^4l(l+1)}{a_0^{-4}})^{1/2}-1 $. 
Thus, we obtain  finite values of  $n$ for   bound states. Now using   $\beta (\lambda),\gamma,A_0,A_1,A_2 $, we can  obtain  the upper and lower bounds of the potential $ V_0$.
\begin{equation} \label{boundV0}
	\frac{8\Delta_{\lambda}}{\alpha}-\frac{\hbar^2}{2m a^2} n^{'2}<V_0<\frac{8\Delta_{\lambda}}{\alpha}+\frac{\hbar^2}{2m a^2}n^{'2} 
\end{equation}
Using this bound it is possible to explicitly obtain the corrected energy of the system as (Appendix C) 
\begin{eqnarray}
E_{nl}(\lambda)   &=&\frac{\hbar^2 l(l+1)\lambda^2}{2m a_0^2} \xi_0  -\frac{\hbar^2}{2m a^2}\left[\frac{\xi_1}{16}  +\frac{\xi_2}{\xi_1}+\frac{m V_0 a^2}{\hbar^2}\right],  
\end{eqnarray}
where  $\xi_0 = 1+12a^2 a_0^{-2}$, $\xi_1 = [(1+192l(l+1)a^4 a_0^{-4})^{1/2}-2n-1]^2$ and $\xi_2 = 4[m a^2 V_0 \hbar^{-2}-4l(l+1)\lambda^2a^3 a_0^{-3}]^2$. 

\begin{figure}
\begin{tabular}{c}
\includegraphics[width=8cm,height=5.5cm]{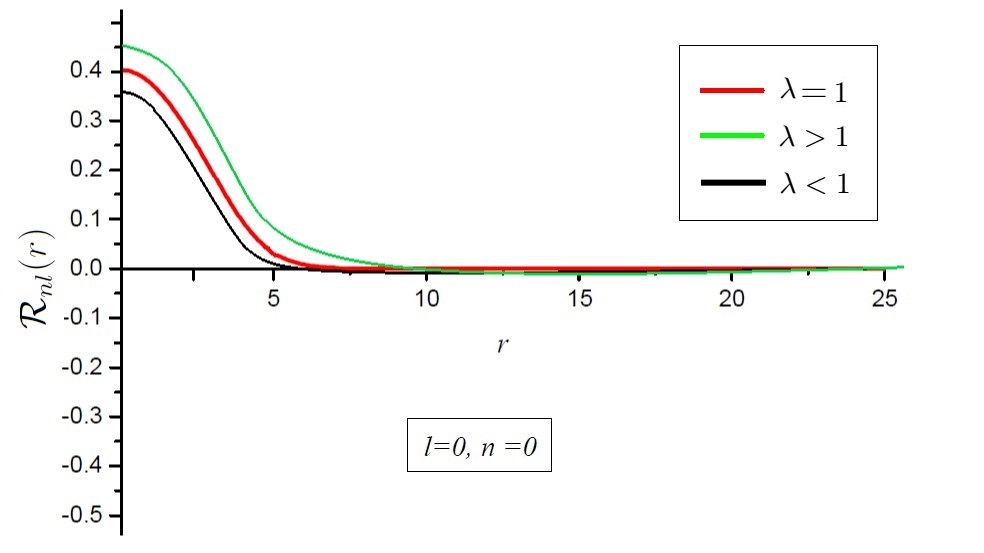} \hspace*{.2cm}\includegraphics[width=.45\linewidth]{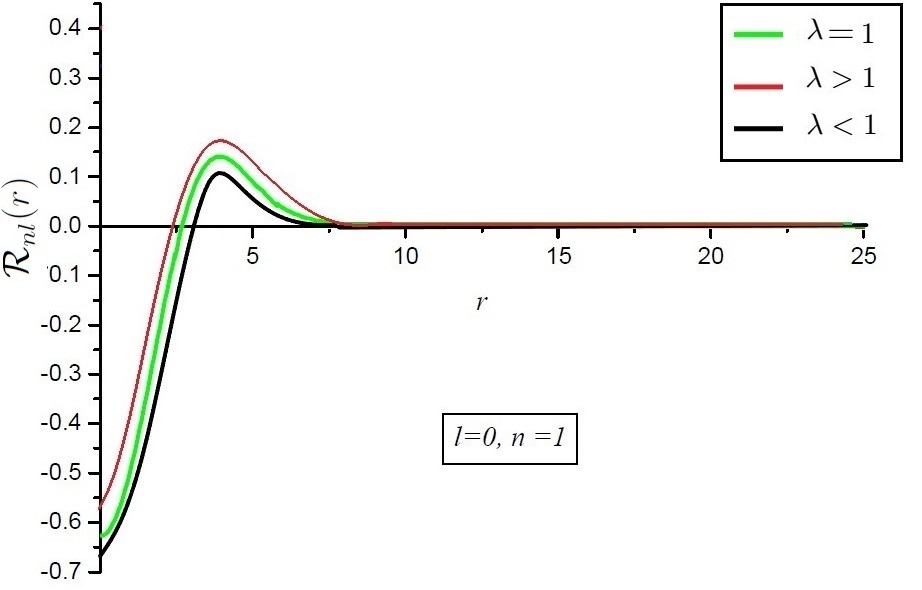}\\ \\
\includegraphics[width=.45\linewidth]{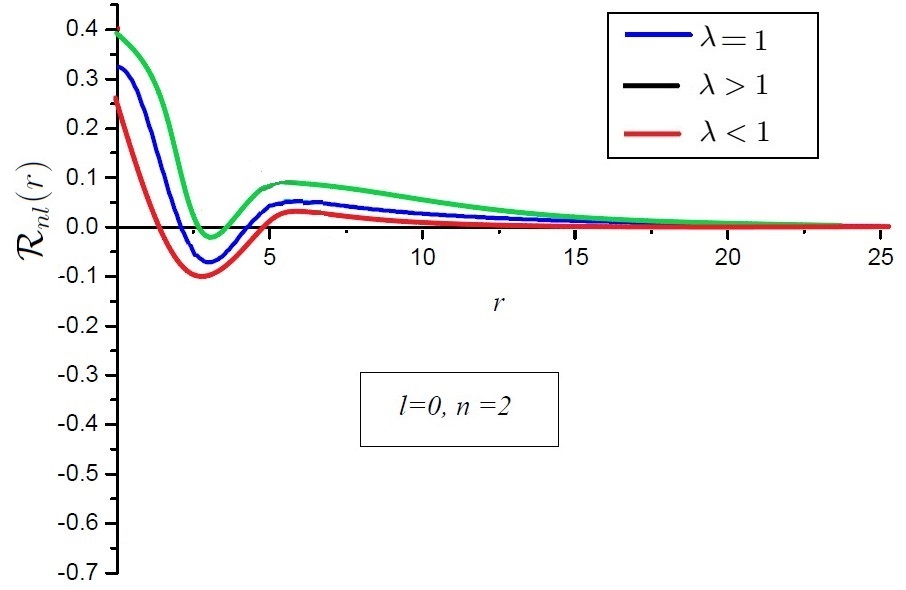} 
\end{tabular}
	\caption{The wave functions for Schr\"{o}dinger's equation,    with Woods-Saxon potential for different values of $l,$ $n$ and $\lambda$. The top figure corresponds to the values $l=0,n=0$.  Second plot corresponds to $l=0,n=1$. The bottom plot corresponds $l=0,n=2$. In each plot, we have plotted the wave functions for $\lambda =1,\geq 1, \leq 1$.}\label{fig_wavefn}
\end{figure}

Now we can write the   wave functions for the  deformed   Woods-Saxon  potential. The general wave function can be expressed as 
\begin{equation}
\psi(z)=\Phi(z)y(z)  \label{w10} 
\end{equation}
where $\Phi(z)=z^{\epsilon(\lambda)}(1-z)^{\zeta}$ with 
$\zeta = \sqrt{\epsilon^2(\lambda)-\beta^2(\lambda)+\gamma^2}$. 
This wave function can be simplified by using the  Rodrigues relation and Jacobi polynomials  (Appendix D). So, using  
$\alpha = 2\epsilon $ and $\beta = 2\sqrt{\epsilon^2(\lambda)\beta^2(\lambda)+\gamma^2} $, we can write the radial wave function  as 
\begin{eqnarray} \label{Rnl_last}
	\mathcal{R}_{nl}(z)=K_{nl}z^{\epsilon(\lambda)}(1-2z)(1-z)^{\sqrt{\epsilon^2(\lambda)-\beta^2(\lambda)+\gamma^2}} P_n^{(2\epsilon(\lambda),2\sqrt{\epsilon^2(\lambda)-\beta^2(\lambda)+\gamma^2}}
\end{eqnarray}
where  $K_{nl}$ is the normalizing constant. 
It may be noted  that the value    $\lambda=1$, corresponds to   $\mathcal{\langle C\rangle}=0$, which produces the    wave functions for  original  Schr\"{o}dinger's equation    with Woods-Saxon  potential. Thus,  $\lambda$  measures the corrections to this equation from the deformation of the Heisenberg algebra. 
We have presented the wave functions for the deformed   Schr\"{o}dinger's equation for the iron nucleus $(^{56}Fe)$ in  Fig. \ref{fig_wavefn}.
The parameters for this system are $a_0=1.285 fm$,  $a=0.65 fm$, $V_0=(40.5+0.13A) MeV =47.78 MeV, R=R_0A^\frac{1}{3}=4.9162 fm, m_{core}=56u,m_n=1.00866u$. We plot the wave functions for $l=0$ and $n=0,1,2$ with $\lambda =1$ (original  Schr\"{o}dinger's equation), and for  $\lambda>1$ and $\lambda<1$. It may be noted that as we are using a general deformation of the Heisenberg algebra  \cite{line12, line14,bb1,bb2,bb3,bb4,bb5,bb6,bb7, line16, line18}, it is possible to take both these values. They correspond to the modification terms in the deformed Heisenberg algebra.

\section{Conclusion}
In this paper, we  have analyzed the correction to the  mean field theory potential for a system of  nucleons from a deformed Heisenberg algebra representing quantum gravitational effects. As the mean field potential for the nucleons can be described by the Woods-Saxon potential, we analyzed such a deformation of  the Schr\"{o}dinger's equation  for the Woods-Saxon potential. We used a Nikiforov-Uvarov method to solve such a deformed  Schr\"{o}dinger's equation. These solutions were numerically analyzed. It was demonstrated that the deformed  Schr\"{o}dinger's equation could be analyzed as an ordinary deformed  Schr\"{o}dinger's equation with an effective potential. This effective potential was expressed as a sum of the deformed and the original Woods-Saxon potentials. We explicitly obtained the corrections to the energy eigenstates of this system. The corrections to the wave function of this system were also obtained. 
It would be interesting to use these results explicitly to analyze various different nuclear systems. Thus, we can use the data from different nuclear processes to analyze such corrections. It could be possible to use such precise nuclear data to constrain quantum gravity theories. Now such a modification of the wave function will modify the dynamics in most physical models described by Woods-Saxon potential. We can use this wave function to calculate the dwell
time in tunneling, which can then be related to the half life of the  nucleus  \cite{k9, k10, k12, k14}. 
This  dwell time is the average time spend by a particle in a given region. 
Now for a nuclear system, with  incident flux $j$, this dwell time  between $z_1$ and $z_2$ is given by  \cite{k9, k10}
\begin{equation}
\tau_D = \frac{1}{j}\int_{z_1}^{z_2} dz |\psi(z)|^2, 
\end{equation}
where $\psi(z)$ is the  wave functions for the      Woods-Saxon  potential.  Now as this deformation of the Woods-Saxon  potential  has modified this wave function, it would modify the dwell time, and this in turn would modify the half life of the  nucleus  \cite{k9, k10, k12, k14}. This modification can be explicitly obtained by using the corrected wave function    given by Eq. (\ref{w10}).
Such modification could be experimentally detected in future experiments. It would be interesting to investigate the effect of such deformation on various nuclear systems, and use the experimental data to constraint the minimal length used to deform such systems. It may be noted that we have used a specific  deformation of the   Heisenberg \cite{line12, line14,bb1,bb2,bb3,bb4,bb5,bb6,bb7, line16, line18} for deforming the Woods-Saxon potential. The results obtained in this paper are consistent with the modifications to the such a nuclear system as produced by several other different approaches. 

Even though we used a deformation of the Heisenberg algebra, which was consistent with various different theories, it would also be interesting to analyze effect of other kind of deformations on the Heisenberg algebra on nuclear potentials. It may be noted that it is also possible to consider more covariant deformations of the Heisenberg algebra \cite{qft2,qft4,qft5,qft6,qft7}.  Such covariant deformations have been used to deform the equations of motion for quantum field theories. It would be interesting to analyze the consequences of such covariant deformation on the mean field theory of nucleons. We expect that such deformation can produce interesting corrections to the energy eigenstates of the system. It is also expected that the decay rates for different nuclear reactions could be modified by such deformations of the system. It would thus be interesting to calculate such corrections to the decay rates for a deformed nuclear system using a deformed Schr\"{o}dinger's equation. It may be noted that the Woods-Saxon potential has been used to model both the entrance channel fusion barrier and the fission barrier of fusion-fission reactions. This was done using the Skyrme energy-density functional approach \cite{fusion1,fusion2}. In that study, the fusion excitation functions of several reactions were investigated. It was observed that the fusion cross-sections could be obtained from the calculated potential for such a system. A statistical model for such a process was used to analyze the decay of the compound nucleus. It was observed that the experimental data was consistent with the calculated based on the Woods-Saxon potential. It would be interesting to investigate the corrections to such a model for the entrance channel fusion barrier and the fission barrier of fusion-fission reactions from quantum gravity. This can again be studied using the corrections to Schr\"{o}dinger's equation by minimal length. It is possible to analyze a direct coupling of laser to nucleus of atoms \cite{laser1,laser2,laser4,laser5}. It has been proposed that such a coupling of lasers to nucleus can be used to produce nuclear fission in those nuclear systems. This process can be investigated using the nuclear double folding potentials. In fact, it is possible to consider the corrections to both these systems from quantum gravity.

\section*{Appendix A}
We will now review Nikiforov-Uvarov method, as it will be used to analyze the deformed 
  Schr\"{o}dinger's equation   with Woods-Saxon  potential.  It may be noted that  Nikiforov-Uvarov method has been used to analyze the original  Woods-Saxon  potential \cite{nu12, nu14}.
  However, we will use it here to analyze a deformation of this  Woods-Saxon  potential.
 So, we  will start with a differential equation of the form 
\begin{equation} \label{NU1}
	\psi''(z) + \frac{\tilde{\tau}(z)}{\sigma(z)} \psi'(z) + \frac{\tilde{\sigma}(z)}{\sigma^2(z)}
\end{equation}
where  $\sigma(z)$ and $\tilde{\sigma}(z)$ are polynomials of at most seond degree. Here we take  $\tilde{\tau}(z)$ as  a first degree polynomial and $\psi(z)$ as  a hyper-geometric function. Now we can  use the transformation,
\begin{equation} \label{NU2}
	\psi(z ) = \Phi(z) y(z)
\end{equation}    
Using this transformation,  we can express Eq. \eqref{NU1} as 
\begin{equation} \label{NU3}
	y''(z) + \left(\frac{ 2\phi'(z)}{\phi(z)} + \frac{\tilde{\tau}(z)}{\sigma(z)} \right) y'(z) + \left( \right)
\end{equation} 
It is possible to define
\begin{eqnarray} \label{NU4}
	\frac{ 2\phi'(z)}{\phi(z)} + \frac{\tilde{\tau}(z)}{\sigma(z)} &=& \frac{\tau(z)}{\sigma(z)}
\\ 
\label{NU5}
	\frac{\phi'(z)}{\phi(z)} &=& \frac{\pi(z)}{\sigma(z)}
\end{eqnarray} 
Now using  polynomials $\tau(z)$ and $\pi(z) = 1/2 [\tau(z) - \tilde{\tau}(z)]$, we can write 
\begin{equation} \label{NU6}
	y''(z) + \frac{\tau(z)}{\sigma(z)} y'(z) + \frac{\tilde{\sigma}(z)}{\sigma^2 (z)} y(z) = 0
\end{equation}
Furthermore,  replacing $\tilde{\sigma}(z) = \omega. \sigma(z)$,  we obtain 
\begin{equation} \label{NU7}
	\sigma(z) y''(z) + \tau(z) y'(z)  + \omega y(z) = 0
\end{equation}
It is possible to obtain  $y(z)$ and $\omega$. This can be done by  using properties of hypergeometric function, and using the relation
\begin{equation} \label{y_rodriguez}
	 y_n (z) = \frac{A_n}{\rho(z)} \frac{d^n}{dz^n} \left(\sigma^n(z) \rho(z)\right)
\end{equation}
Here  $\rho(z)$ satisfies the following equation \begin{equation}
	(\sigma(z)\rho(z))'  =\tau(z) \rho(z)
\end{equation}
The function $\pi(z)$ and the parameter $\omega$ can be expressed as 
\begin{equation} 
\pi(z) = \frac{\sigma'-\tilde{\tau}}{2} \pm \sqrt{\left(\frac{\sigma'-\tilde{\tau}}{2}\right)^2 - \tilde{\sigma} + \mathcal{P} \sigma}
\end{equation}

\begin{equation} \label{omega_def}
\omega=\mathcal{P} +\pi'(z)  
\end{equation}
The  eigenvalues of Eq. \eqref{NU1} are given by 
\begin{equation} \label{NU_eigenval}
	\omega_n = -n \tau'(z) - \frac{n(n-1)}{2} \sigma''(z)
\end{equation}

 \section*{Appendix B}
We analyze the   total  effective potential, which is given by a sum of the original Woods-Saxon potential $V_{WS} (r)$ and the deformation of the potential  $V_l (r)$  
\begin{eqnarray} 
	V_{eff} = V_l(r) + V_{WS}, && 
	 V_l(r) = \frac{\hbar^2 l(l+1)}{2mr^2} \lambda^2
\nonumber 	\\
	V_{WS} = \frac{V_0}{1+ exp(\frac{r-a_0}{a})}. && 
\end{eqnarray}
To analyze this potential we will use   two new variables  
\begin{eqnarray}
x = \frac{r- a_0}{a_0}, &&
\alpha = \frac{a_0}{a}.
\end{eqnarray}
 We can now solve the effective potential using   Pekeris approximation 
\begin{eqnarray}
V_{WS} &=& - \frac{V_0 }{1+ exp(\alpha x)}, ~\label{WS_x}
\\
V_l(x)  &=&  \frac{\hbar^2 l(l+1)}{2m a_0^2 (1+x)^2} \lambda^2 = \frac{\Delta_\lambda}{(1+x)^2} ~\label{Vl_x}
\end{eqnarray}
We will replace $V_l(x)$ by $V^*_l (x)$, and write 
\begin{equation} \label{vl_pek_1}
	V^*_l (x)= \Delta_\lambda \left[A_0 + \frac{A_1}{1+ exp(\alpha x)} + \frac{A_2}{(1+ exp(\alpha x))^2} \right]
\end{equation}
To obtain  the coefficients $ A_0,  A_1, A_2$, we use taylor expansion about   $x =0$  
\begin{eqnarray} \label{taylor}
	V^*_l (x) &=& \Delta_\lambda \left[ \left(A_0 + \frac{A_1}{2} +\frac{A_2}{4}\right) - \frac{\alpha}{4} \left( A_1+ A_2 \right) x \right.  \nonumber \\  && \left. + \frac{\alpha^2}{16} A_2 x^2 + \frac{\alpha^3}{48} \left(A_1 +A_2\right) x^3 \right]
\end{eqnarray}
Now using the   expansion of $V_l(x)$ as 
$ V_l(x)= \Delta_\lambda \left[1- 2x +3x^2 -4x^3 \right]$, we can write  the coefficients as  $
	A_0 = 1- {4}{\alpha}^{-1} + {12}{\alpha^{-2}},   A_1 = {8}{\alpha}^{-1} -{48}\alpha^{-2}, A_2 = {48}\alpha^{-2}
$.  
Thus, the effective potential can be written as 
\begin{eqnarray}\label{eff_pek}
	V^*_{eff}(x) &=& V^*_l(x) + V_{WS} (x)\nonumber \\ &=& \Delta_\lambda A_0 - \frac{V_0 - \Delta_\lambda A_1}{1+ exp(\alpha x)} + \frac{\Delta_\lambda A_2}{[1+ exp(\alpha x)]^2}
\end{eqnarray}

\section*{Appendix C}
We can solve the  deformed Schr\"{o}dinger's equation   by   Nikiforov-Uvarov method \cite{nu12, nu14}.  The deformed  Schr\"{o}dinger's equation is shown in Eq. \eqref{GUP_SE4}.
Now using  $
	\tilde{\tau}(z) = 1-2z, \sigma(z) = z(1-z),  \tilde{\sigma}(z) = -\epsilon^2(\lambda) + \beta^2(\lambda) z  - \gamma^2 z^2 $, we can write 
the  function $\pi(z)$ of   Nikiforov-Uvarov method as \cite{nu12, nu14} 
\begin{equation}
	\pi(z) =\pm \sqrt{f(z)} = \pm \sqrt{\epsilon^2(\lambda)  + [\mathcal{P} - \beta^2(\lambda)]z  - [\mathcal{P} - \gamma^2]z^2} 
\end{equation}
The parameter $\mathcal{P}$ can be found by using the solution of the equation  $f(z) =0$. 
Thus, we can write the  solutions to this equation as $	 \mathcal{P} =  \pm [ \epsilon(\lambda) -  A(\lambda)] z - \epsilon(\lambda)$ and  $	\mathcal{P} =   \pm [ \epsilon(\lambda) + A(\lambda)  ] z - \epsilon(\lambda)$, with $A(\lambda) = \sqrt{\epsilon^2 (\lambda)- \beta^2(\lambda) +\gamma^2} $. 
It may be noted that  $\tau(z)$ should reduce with $z$  and its roots are required to be in the interval $(0,1)$. Therefore the appropriate form  of $\pi(z)$ and $\tau(z)$ can be expressed as 
\begin{align}
\pi(z)=\epsilon(\lambda)-\left[\epsilon(\lambda)+\sqrt{\epsilon^2(\lambda)-\beta^2(\lambda)+\gamma^2}\right]z ~\label{pi_sol}
\\
\tau(z)=1+2\epsilon-2\left[1+\epsilon+\sqrt{\epsilon^2(\lambda)-\beta^2(\lambda)+\gamma^2}\right]z  ~\label{tau_sol}
\end{align}
Now  we can write the parameter $\omega$ for this solution as   
\begin{eqnarray}\label{omega_sol}
	\omega &=&\beta^2(\lambda)-2\epsilon^2(\lambda)-2\epsilon(\lambda)\sqrt{\epsilon^2(\lambda)-\beta^2(\lambda)+\gamma^2}-\epsilon(\lambda)\nonumber \\ && +\sqrt{\epsilon^2(\lambda)-\beta^2(\lambda)+\gamma^2} 
\end{eqnarray}
We can also  write the eigenvalues  $\omega_n$ as 
\begin{equation} \label{omega_n_sol}
	\omega_n =2\left[\epsilon(\lambda)+\sqrt{\epsilon^2(\lambda)-\beta^2(\lambda)+\gamma^2}\right]n+n(n+1)
\end{equation}

It may be noted that this deformed  Woods-Saxon  potential is finite.  Condition for the bound states to exist in this deformed Woods-Saxon potential is given by 

\begin{equation} \label{bound}
-n^{'2}<\left[\beta^2(\lambda)-\gamma^2\right]<n^{'2}, 
\end{equation} 
$n^{'}= -n + {\sqrt{1+4\gamma^2}-1}/{2}
$. 
The bound states only exist for $n^{'} > 0$. Using the value of $\gamma $, we can   
 write this bound as 
\begin{equation} \label{con_n2}
0 \le n\le \frac{\sqrt{1+\frac{192a^4l(l+1)}{a_0^4}}-1}{2}
\end{equation}
Thus, we get finite values of  $n$ for those  bound states. Now using   $\beta (\lambda),\gamma,A_0,A_1,A_2 $, we can  obtain  the upper and lower bounds of the potential $ V_0$.
\begin{equation} 
	\frac{8\Delta_{\lambda}}{\alpha}-\frac{\hbar^2}{2m a^2} n^{'2}<V_0<\frac{8\Delta_{\lambda}}{\alpha}+\frac{\hbar^2}{2m a^2}n^{'2} 
\end{equation}

Furthermore,  substituting the values of $\Delta_{\lambda} ,\alpha,n^{'},\gamma^2$, we obtain 
\begin{equation}
    V_{0min} < V < V_{0max}
\end{equation}
where 
\begin{eqnarray}
V_{0min}=\frac{4\hbar^2 a l(l+1)\lambda^2}{m a_0^3}-\frac{\hbar^2}{8m a^2}\left[\sqrt{1+\frac{192a^4l(l+1)}{a_0^4}}-2n-1\right]^2 ~\label{V0min_def}
\\
V_{0max}=\frac{4\hbar^2 a l(l+1)\lambda^2}{m a_0^3}+\frac{\hbar^2}{8m a^2}\left[\sqrt{1+\frac{192a^4l(l+1)}{a_0^4}}-2n-1\right]^2 ~\label{V0max_def}
\end{eqnarray}
We can thus write the  energy eigenvalues for the deformed Woods-Saxon potential as 
\begin{equation} \label{Enl1}
	E_{nl}=\Delta_\lambda A_0-(V_0-\Delta_\lambda A_1)\left[\frac{n^{'2}+\beta^2(\lambda)-\gamma^2}{2\beta(\lambda) n{'}}\right]^2
\end{equation}
Using  the values of $\Delta_\lambda,A_0,A_1,A_2,n^{'},\beta(\lambda)$ and $\gamma$, we  can  obtain $E_{nl}(\lambda)$  

\begin{eqnarray}
E_{nl}(\lambda)   &=&\frac{\hbar^2 l(l+1)(1-\mathcal{\langle C\rangle})^2}{2m a_0^2}\left(1+\frac{12a^2}{a_0^2}\right) \nonumber \\ && -\frac{\hbar^2}{2m a^2}\left[\frac{\left[\sqrt{1+\frac{192a^4l(l+1)}{a_0^4}}-2n-1\right]^2}{16} \right. \\ \nonumber && \left.   +\frac{4\left[\frac{m a^2 V_0}{\hbar^2}-\frac{4l(l+1)(1-\mathcal{\langle C\rangle})^2a^3}{a_0^3}\right]^2}{\left[\sqrt{1+\frac{192l(l+1)a^4}{a_0^4}}-2n-1\right]^2}+\frac{m V_0 a^2}{\hbar^2}\right] ~\label{eigenval_final}
\end{eqnarray}

\section*{Appendix D}
Now we can write the   wave functions for the  deformed   Woods-Saxon  potential. The general wave function can be expressed as 
\begin{equation}
\psi(z)=\Phi(z)y(z)  
\end{equation}
where
\begin{equation}
\Phi(z)=z^{\epsilon(\lambda)}(1-z)^{\sqrt{\epsilon^2(\lambda)-\beta^2(\lambda)+\gamma^2}}
\end{equation}
Now using    Rodrigues relation, we can write 
\begin{eqnarray} \label{y_n}
y_n(z)&=&\frac{1}{n!}z^{-2\epsilon}(1-z)^{-2\sqrt{\epsilon^2(\lambda)+\beta^2(\lambda)+\gamma^2}} \nonumber \\ && \times \frac{d^n}{dz^n}\left[z^{n+2\epsilon}(1-z)^{n+2\sqrt{\epsilon^2(\lambda)\beta^2(\lambda)+\gamma^2}}\right]
\end{eqnarray}
Now writing this equation using the   Jacobi polynomials as 
\begin{equation} \label{y_jacobi}
	y_n(z) = \frac{1}{n!}z^{-\alpha}(1-z)^{-\beta}\frac{d^n}{dz^n}\left[z^{n+\alpha}(1-z)^{n+\beta}\right] = P_n^{\left(\alpha,\beta\right)}(1-2z)
\end{equation}
where we have used 
$
\alpha = 2\epsilon $ and $
\beta = 2\sqrt{\epsilon^2(\lambda)\beta^2(\lambda)+\gamma^2} $. 
So, the radial wave function can  be written as,
\begin{eqnarray} 
	\mathcal{R}_{nl}(z)=K_{nl}z^{\epsilon(\lambda)}(1-2z)(1-z)^{\sqrt{\epsilon^2(\lambda)-\beta^2(\lambda)+\gamma^2}} P_n^{(2\epsilon(\lambda),2\sqrt{\epsilon^2(\lambda)-\beta^2(\lambda)+\gamma^2})}
\end{eqnarray}
Here  $K_{nl}$ is a normalising constant and can be found by normalizing conditions
\begin{equation} \int_{0}^{\infty} \left[\mathcal{R}_{nl}(z)\right]^2  dz =1
\end{equation}
Thus, we have obtained the corrections to the wave function from the deformation of the Heisenberg algebra.

\end{document}